\documentclass[letter,
               keeplastbox,   
               ]{jacow}
\usepackage{pdfpages,multirow,ragged2e} %
\makeatletter%
	\ifboolexpr{bool{xetex}}
	 {\renewcommand{\Gin@extensions}{.pdf,%
	                    .png,.jpg,.bmp,.pict,.tif,.psd,.mac,.sga,.tga,.gif,%
	                    .eps,.ps,%
	                    }}{}
\makeatother

\ifboolexpr{bool{xetex} or bool{luatex}} 
 {}                                      
 {\usepackage[utf8]{inputenc}}           

\usepackage[USenglish]{babel}

\ifboolexpr{bool{jacowbiblatex}}%
 {%
  \addbibresource{jacow-test.bib}
  \addbibresource{biblatex-examples.bib}
 }{}
\begin{document}

\title{Digital LLRF system for TRIUMF ISIS buncher}

\author{Xiaoliang Fu\thanks{xfu@triumf.ca}, Ken Fong, Qiwen Zheng, Thomas         Au, Ramona Leewe, \\ 
        TRIUMF, Vancouver, BC, V6T 2A3, Canada 
		}
	
\maketitle

\begin{abstract}
The ISIS buncher system at TRIUMF operates at frequencies of 23MHz, 46MHz, and 4.6MHz. The 23MHz and 46MHz signals drive two buncher cavities, while the 4.6MHz signal drives the 5:1 selector. The previous analog-digital hybrid system has been replaced with a new digital LLRF system due to occasional drifts in the setpoints of the control loops during operation. The reference signal for the LLRF system is obtained from the pickup signal of the cyclotron's cavity, ensuring that all frequencies are synchronized with it. In the event of a spark occurring in the cyclotron's cavity, the LLRF system may lose its reference signal. To address this, a phase-locked loop with a track and hold function is designed to maintain the frequency when the reference signal is absent. The 4.6MHz frequency is derived by dividing the 23MHz reference signal frequency by 5. Designing the divide-by-5 circuitry posed specific challenges in a binary system. The LLRF system is built upon TRIUMF's versatile digital LLRF hardware system, with firmware optimized specifically for the ISIS buncher system. This paper will delve into the details of the hardware and firmware.
\end{abstract}

\section{Introduction}
\begin{figure}[!htb]
\includegraphics[width=\hsize]{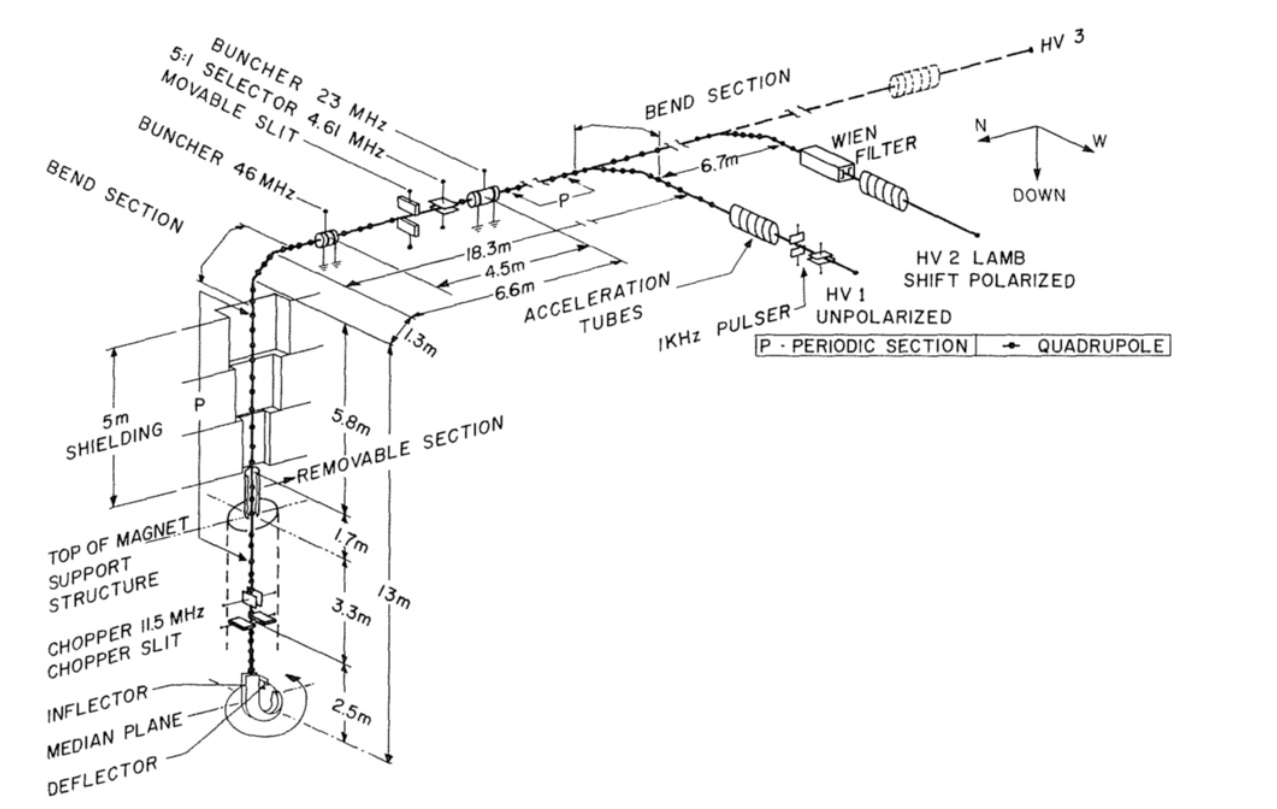}
\caption{Injection line layout.~\cite{b1}}
\label{fig1}
\end{figure}

The Ion Source and Injection System (ISIS) of TRIUMF is a 40 m long electrostatic beamline between the ion source and the 520 Mev cyclotron, as shown in Fig.~\ref{fig1}. The $H^{-}$ ions are produced by two external ion sources and are transported at ~300 KeV along the injection beamline to the cyclotron. Two double-gap sinusoidal bunchers are located in the middle of the beamline to enhance the cyclotron acceptance of the DC beam.The initial buncher operates at the cyclotron RF frequency (23.06MHz), and the other buncher, located 4.54 m downstream, operates at the second harmonic (46.12MHz). The voltage and phase of each buncher are independently adjusted to optimize the beam accepted by the cyclotron. Additionally, a 5:1 selector, which operates at 1/5 of the cyclotron RF frequency (4.62MHz), is located between the two bunchers to increase the time between beam bursts by selecting every fifth bunch into the cyclotron. The LLRF system accepts the reference signal from the cyclotron RF system. Due to the cyclotron's resonator hot-arm vibration, the RF drive phase and the resonator's voltage phase is different. The beam phase is matched to the voltage phase, not to the input phase. Therefore, the ISIS buncher's phase is synchronize to the resonator's voltage phase. A pick up signal from the resonator is used as the reference signal for the LLRF system for ISIS bunchers.The cyclotron starts up from the self-excited mode, then switch to the generator driven mode for normal operation. During the self-excited mode, the frequency of the RF system is changing according to the self-excited loop delay, then in the generator driven mode, the RF frequency is fixed to the DDS frequency in the cyclotron's LLRF system. The generator driven frequency can also be changed by the operator during operation. 

\section{System design}
\begin{figure}[!htb]
\includegraphics[width=\hsize]{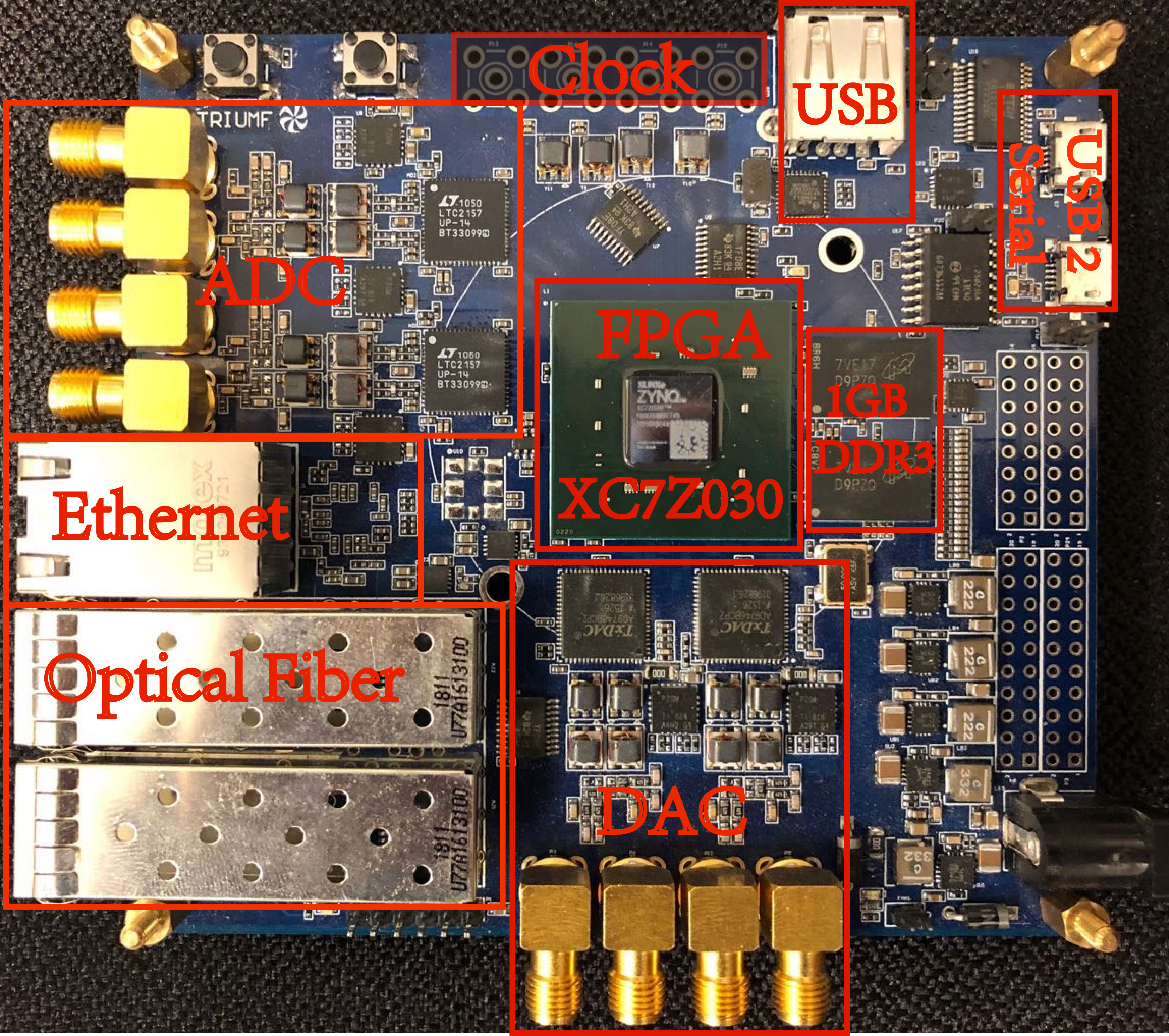}
\caption{Hardware of Digital LLRF system of TRIUMF.}
\label{fig2}
\end{figure}

The digital LLRF system~\cite{b2} at TRIUMF is based on ZYNQ FPGA, equipped with high-speed ADC and DAC, as shown in Fig.~\ref{fig2}. The design concept aims to create a versatile digital LLRF system that can be utilized throughout TRIUMF's accelerator system. It incorporates a large-capacity FPGA and a powerful CPU capable of running an embedded Linux system. Over the past three years, the first-generation digital LLRF system has been successfully deployed in ISAC I, e-Linac, and ISIS bunchers(two harmonics) systems. During the 2023 winter shutdown, the LLRF group has upgraded the control system for DTL4/5, ISIS bunchers(two harmonics with 5:1 selector), and HEBT11/35 systems with the digital LLRF hardware system.

Each of TRIUMF's digital LLRF board supports a maximum of three channels for RF amplitude and phase control. The reference input signal is utilized to phase-lock the LLRF system to an existing RF system. To achieve these functions, the LLRF system is optimized in both FPGA firmware and Linux software. Leveraging the FPGA's ample logic capacity, various high-speed interfaces for ADC/DAC, RF demodulation and modulation circuits, phase-locked loop circuits, PID controllers for amplitude and phase, and tuner controllers are implemented inside the FPGA using HDL code. These FPGA circuits serve as peripherals to the CPU. The ARM CPU controls the system by modifying the parameters of the hardware within the FPGA. The Linux operating system runs on the dual-core ARM CPUs and manages all the CPU peripherals. Hardware parameters, controlled by registers, are presented as Linux character devices, allowing the mid-layer driver to easily control them. The LLRF system adopts a Python file as its mid-layer driver. This Python file contains a USB command set, a function library, an EPICS command set, and firmware drivers. It hides all the hardware details from users. 

\subsection{FPGA firmware}

\begin{figure*}[!htb]
\centering
\includegraphics[width=0.7\hsize]{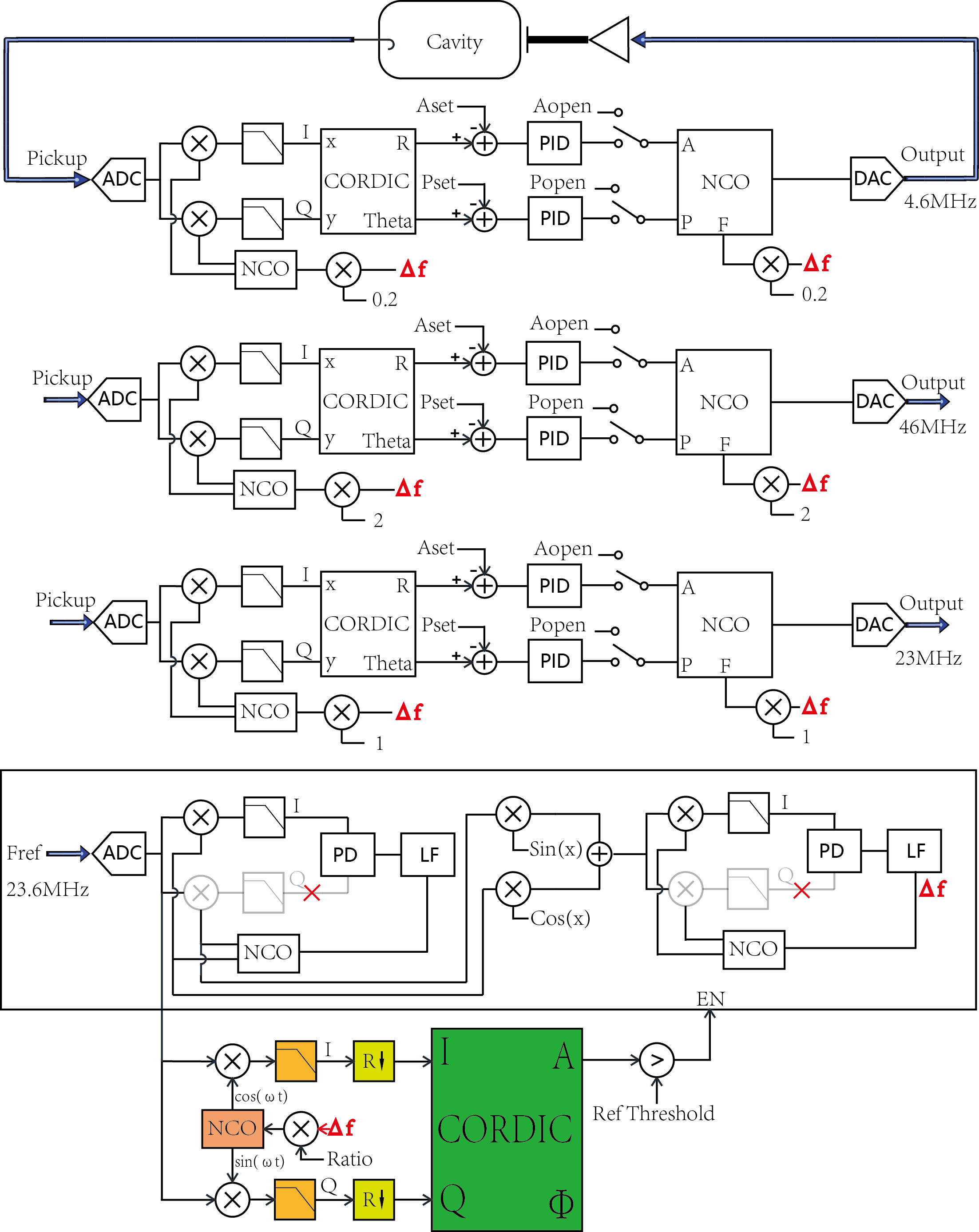}
\caption{Block diagram of ISIS buncher LLRF firmware.}
\label{fig3}
\end{figure*}

The FPGA firmware is designed with the fundamental concept of creating a versatile hardware platform capable of demodulating and modulating any frequency within the ADC and DAC bandwidth at runtime. This eliminates the need to change the hardware every time the frequency requirements change. Taking inspiration from software-defined radio principles, the FPGA firmware utilizes a Digital Down-Conversion (DDC) module for demodulation and a Direct Digital Frequency Synthesis(DDS) module for modulation. By dynamically adjusting the center frequency of the modulation and demodulation modules through software control, the system achieves flexibility and reusability. This approach allows the hardware to adapt to different operating frequencies without requiring physical modifications. As a result, the system becomes highly flexible, enabling efficient frequency handling and reducing the need for hardware changes in diverse applications. The system diagram is shown in Fig.~\ref{fig3}.

\subsubsection{Digital Phase-locked Loop and Global Phase Shifter}

The ISIS bunchers LLRF system operates in driven mode, and the output RF signal is generated by the DDS that is not phase-locked to the reference signal. In order to address the phase synchronization issue in the ISIS LLRF system, a digital phase-locked loop (PLL) is incorporated. In the first version firmware, a Costas loop, widely used in communication systems, was employed for the digital PLL. The Costas loop allowed the frequency of the Numerically Controlled Oscillators (NCOs) to be locked to the reference signal. Additionally, the harmonic frequencies could be locked to the reference signal by adjusting the frequency tuning words of the NCOs. However, an issue arose in the first version PLL due to the presence of two stable operating points in the range of [0, 359] degrees in the phase detector. While this was generally not problematic in most applications, it became a critical issue in the ISIS buncher project due to the fact that the reference signal comes from the resonator. If a spark occurred in the resonator, the RF drive signal would be turned off by the spark protection circuit, resulting in the loss of the pickup signal. In such cases, the LLRF system would lose its reference, and it was crucial for the system to maintain the correct frequency once the reference signal was restored. However, there was a possibility that the phase-locked loop would lock onto the other operating point of the phase detector, leading to a 180 degree phase error between the reference signal and the LLRF system output. Consequently, the beam line would immediately lose the beam. This phenomenon was observed multiple times during the beam operation. To resolve this issue, the phase detector within the phase-locked loop had to be redesigned to have only one operating point in the range of [0, 359] degrees. This involved removing the previous quadrature phase detector from the design and introducing a multiplier phase detector to address the problem effectively.
Assume that the reference signal is:
\begin{equation}
 \label{eq1}
x(t)=A\cos(\omega_{c} t)
\end{equation}
And the output of the local NCO is:
\begin{align}
Q_{0} &=\sin(\omega_{0}t+\phi(t)) \label{eq2}
\end{align}
The mixing results of reference signal and NCO signal are:
\begin{align}
Q_{o} &= A\cos(\omega_{c} t)\cdot \sin(\omega_{0}t+\phi(t)) \label{eq3}
\end{align}

After a the low pass filters, the higher frequency $\omega_{c}+\omega_{0}$ is attenuated, which can be ignored:
\begin{equation}
Q_{o}=A/2\cdot\sin[(\omega_{c}-\omega_{0})t-\phi(t)] \label{eq4}
\end{equation}
After the phase detector, which is also a multiplier, the result is:
\begin{equation}
 \label{eq5}
P_{e}(t)=A/8\sin((\omega_{c}-\omega_{0})t-\phi(t))
\end{equation}
For the quadrature phase detector, the phase error is:
\begin{equation}
 \label{eq6}
P_{eq}(t)=K/8\sin(2(\omega_{c}-\omega_{0})t-2\phi(t))
\end{equation}

If a multiplier phase detector is used instead, the phase error is:
\begin{equation}
 \label{eq7}
P_{em}(t)=K/8\sin((\omega_{c}-\omega_{0})t-\phi(t))
\end{equation}

Based on Eq.~\eqref{eq6}, the phase error for the quadrature phase detector $P_{eq}(t)$ is a function of $2(\omega_{c}-\omega_{0})t-2\phi(t)$. Therefore, there will be two stable operation points in [0,359] for feedback control. The phase-locked loop can operate at any operation point, but the output of the LLRF will have a 180 degree phase error that is unacceptable to the beam. Eq.~\eqref{eq7} shows that there is only one operation point in [0,359], the phase-locked loop will always lock to one operation point when the reference signal go back and forth. The loop filter plays a crucial role in the digital phase-locked loop (PLL), and an active Proportional-Integral (PI) filter is employed as the loop filter instead of a Finite Impulse Response (FIR) filter due to its lower delay. The PI parameters of the loop filter are optimized to reduce controller overshoot and minimize system phase noise. The loop filter calculates the frequency correction value, which is subsequently applied to the NCO to correct its frequency. This frequency correction value serves as the foundational reference for the system. By adjusting the ratio of the frequency correction value, it becomes possible to program and phase-lock the frequency of each channel to the reference signal. As a unique scenario, the 5:1 selector operates at 1/5 of the cyclotron RF frequency, resulting in a fractional frequency ratio. Obtaining an exact fractional value in the FPGA using fixed-point numbers is impractical. As the fixed-point bit width increases, the frequency error decreases, although it cannot reach absolute zero. However, even a minor frequency error can contribute to a gradual phase drift between the reference signal and the 5:1 selector's output during extended periods of operation. This challenge is addressed by periodically adjusting the frequency ratio number to maintain overall phase stability. Switching the frequency ratio between 0.2000000000262 and 0.2000000000261 will keep the phase stable. Since the phase stability requirements for the 5:1 selector is 10 degree, the phase control performance is more than enough. The foundational frequency and the ratio of each channel can be changed by the software at run-time.

When the resonator sparks, the reference signal will be gone and the PLL has nothing to track. The PLL has to be disabled when the reference signal is not present. A monitoring module is incorporated to measure the amplitude of the reference signal. When the reference signal is lost during the resonator sparking, the monitoring module disables all the modules in the phase-locked loop and holds their current values. The LLRF system then operates at a fixed frequency that is few hundreds Hz away from the moment that the reference signal is lost. After the reference signal came back, the monitor module enable the PLL again and the PLL starts to track the reference signal. The threshold of the monitoring module can be configured by the software. It is important to set the threshold value properly to make sure the noise on the pickup cable won't trigger the PLL to track to noise signals.

The ISIS buncher LLRF system has implemented a global phase shifter inside the FPGA, enabling software global phase control during run-time. The global phase shifter is inserted between the first and the second phase-locked loop. Since system is phase-locked to the second phase-locked loop, the adjustment of the system's global phase will effect all the RF channels. The global phase shifter is based on an I/Q modulator and works open loop. The ARM CPU accepts the global phase setting point $\phi$ from the local control computer, calculates the fixed-point $sin(\phi)$ and $cos(\phi)$ value, and controls the phase shifter by AXI-GPIOs.

\subsubsection{Amplitude and Phase control} 

For a single frequency system, assume the the cavity pick up signal is~\cite{b2}:
\begin{equation}
 \label{eq8}
u_{0}(t)=A_{0}[1+f(t)]\cos[\omega t+\phi^{'}(t)]
\end{equation}
where $f(t)$ is the amplitude modulation signal of the cavity. The output signal of the local NCO in the DDC module is:
\begin{align}
I&=\cos(\omega_{1}t)  \\
Q&=\sin(\omega_{1}t)
\end{align}

The mixing results of cavity signal and NCO signal can be written as:
\begin{equation}
 \label{eq11}
\begin{cases}
I_{1}  &=A_{0}[1+f(t)]/2\{ \cos[(\omega+\omega_{1}) t \\
& \quad +\phi^{'}(t)]+\cos[(\omega-\omega_{1})t+\phi^{'}(t)]\} \\
Q_{1} &=A_{0}[1+f(t)]/2\{ \sin[(\omega+\omega_{1}) t \\
& \quad +\phi^{'}(t)]+\sin[(\omega-\omega_{1})t+\phi^{'}(t)]\}
\end{cases}
\end{equation}
After a the low pass filters, the higher frequency $\omega+\omega_{1}$ is attenuated, which can be ignored:

\begin{equation}
 \label{eq12}
\begin{cases}
I_{1}  &=A_{0}[1+f(t)]/2\cdot \cos[(\omega-\omega_{1})t+\phi^{'}(t)] \\
Q_{1} &=A_{0}[1+f(t)]/2\cdot \sin[(\omega-\omega_{1})t+\phi^{'}(t)]
\end{cases}
\end{equation}

The amplitude can be calculated by:
\begin{equation}
 \label{eq13}
U_{1}(t)=\sqrt{I_{1}^{2}(t)+Q_{1}^{2}(t)}=\frac{A_{0[1+f(t)]}}{2}
\end{equation}
Eq.~\eqref{eq13} indicates that the amplitude is independent from phase, frequency, and their modulation.

The phase of the cavity pickup signal is:
\begin{align}
 \label{eq14}
\phi^{'}(t)&=\arctan \frac{Q_{1}}{I_{1}}
\end{align}

Eq.~\eqref{eq14} demonstrates that the phase error remains unaffected by variations in amplitude, frequency, and their modulation. As a result, the phase control is independent of the phase-locked loop. Following the demodulator's calculation of the amplitude error and phase error, a closed-loop amplitude and phase control is achieved through a PID controller.

\subsection{Software based on Linux OS}

The ISIS buncher LLRF system adopts Debian 11 as the Linux file system. The source code of Debian 11 is compiled and customized specifically for LLRF system. Python 3 is chosen as the default Python environment, and GCC 10.2.1 is also available for users to compile source code such as EPICS locally. The ISIS buncher LLRF system runs a Jupyter Notebook server by default, allowing users to access it through the internet. The default port for Jupyter Notebook is set to 192.168.34.8:8888. With any web browser, users can control the LLRF system using Python code. Popular Python libraries such as NumPy, SciPy, pandas, matplotlib, and scikit-learn are also installed in the system. It is possible to train a machine learning model on a PC and then run the trained model on the LLRF system, opening up new possibilities. 

The ARM CPU controls the FPGA hardware by reading and writing through the AXI-GPIOs. Linux device drivers are developed for the AXI-GPIOs, which are compiled into the Linux kernel and loaded during OS booting. However, since the device driver is close to the hardware level, it is not very convenient for users to directly call the device driver. A mid-layer driver written in Python is used to wrap the hardware details, making the system more user-friendly.

The ISIS buncher LLRF system is controlled by a local control computer through the USB port. The LLRF system is working as a USB HID device under windows operating system. A customized 48 bytes USB command is used to communicate between the LLRF system and the local control computer.

\section{Test and Commissioning}

\begin{figure}[!htb]
\includegraphics[width=.9\hsize]{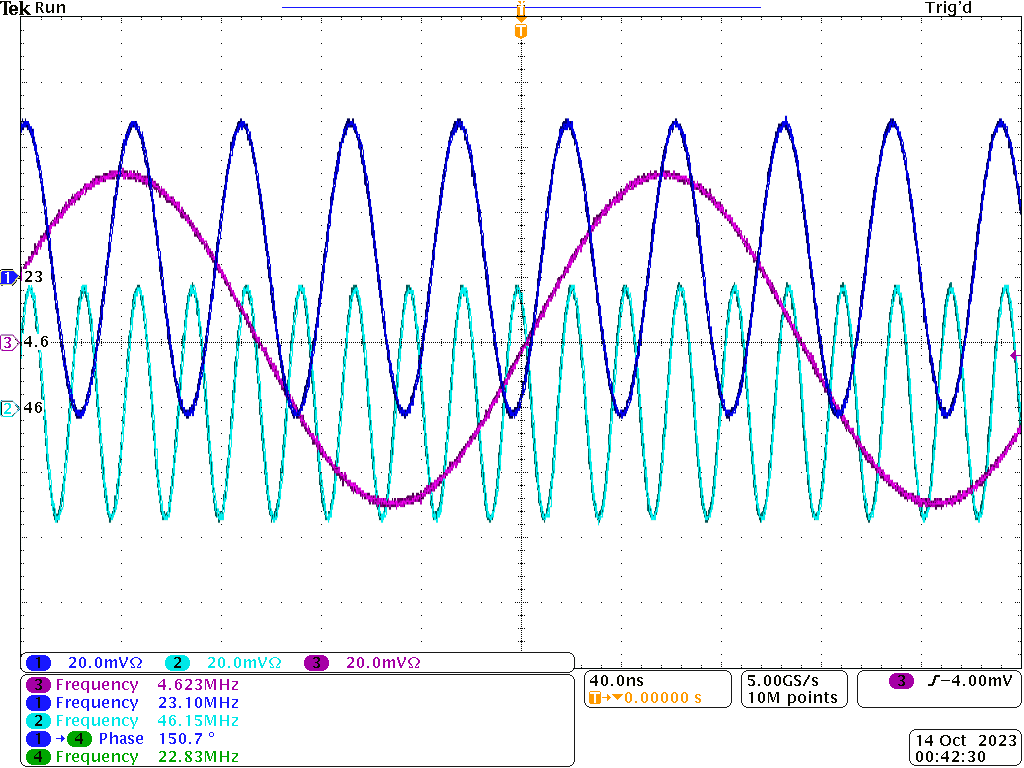}
\caption{Time domain test of the system.}
\label{fig4}
\end{figure}

\begin{figure}[!htb]
\includegraphics[width=.9\hsize]{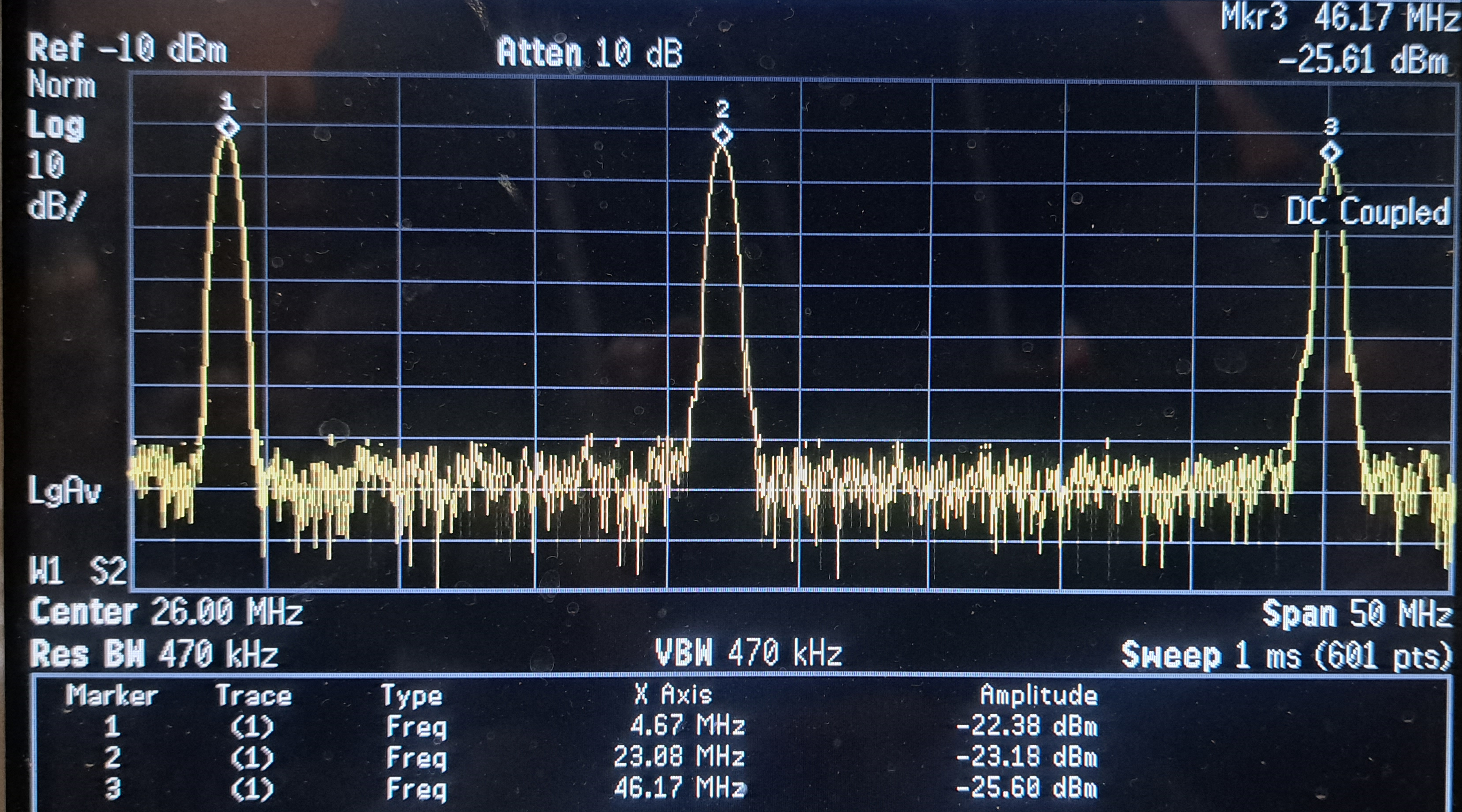}
\caption{Frequency domain test of the system.}
\label{fig5}
\end{figure}

The performance of the LLRF system has been tested on the test bench. The system has been running for 24 hours to check if there is any phase drift between the 4.6MHz and 23MHz. The test results are positive, shown in Fig.~\ref{fig4}. The three outputs of the system are connected to a RF power combiner to check the spectrum, shown in Fig.~\ref{fig5}. The test result indicates that the there is no side-bands around the three signals. After the bench test, the LLRF system is installed on the ISIS buncher system and tested with beam. The commissioning with beam is successful.

\section{Conclusion}

A digital LLRF system has been build for ISIS buncher. The new LLRF system is built based on the Linux operating system and serves as an all-programmable platform. During the 2023 winter shutdown, the LLRF group successfully installed the digital LLRF control systems to the ISIS buncher system. The initial test results for the system with beam have been positive, indicating the improved performance and functionality of the LLRF system. The digital LLRF system for ISIS buncher demonstrated greater abilities, flexibility, and ease of use, providing a significant advancement for the LLRF group and facilitating the control of RF systems at TRIUMF.

%
%
\ifboolexpr{bool{jacowbiblatex}}%
	{\printbibliography}%
	{%
	
	
} 

%
%


\end{document}